# Prevalence of web trackers on hospital websites in Illinois.

Robert Robinson
Department of Internal Medicine
SIU Medicine
Springfield, IL, USA
rrobinson@siumed.edu

## Abstract

Web tracking technologies are pervasive and operated by a few large technology companies. This technology, and the use of the collected data has been implicated in influencing elections, fake news, discrimination, and even health decisions.  Little is known about how this technology is deployed on hospital or other health related websites.

The websites of the 210 public hospitals in the state of Illinois, USA were evaluated with a web tracker identification tool.  Web trackers were identified on 94% of hospital webs sites, with an average of 3.5 trackers on the websites of general hospitals.  The websites of smaller critical access hospitals used an average of 2 web trackers.  The most common web tracker identified was Google Analytics, found on 74% of Illinois hospital websites.  Of the web trackers discovered, 88% were operated by Google and 26% by Facebook.

In light of revelations about how web browsing profiles have been used and misused, search bubbles, and the potential for algorithmic discrimination hospital leadership and policy makers must carefully consider if it is appropriate to use third party tracking technology on hospital web sites.

## Introduction

Web tracking technologies can be found on the majority of all high traffic web sites [Englehardt and Narayanan 2016].  These technologies are used to measure and enhance the impact of advertising and to create web visitor profiles of individual internet users.  This behavioral data feeds the advertising ecosystem that generates revenue for technology companies such as Google and Facebook [Dunn 2017].

This advertising ecosystem has been shown to enforce discrimination by hiding ads for high paying jobs from visitors identified as female [Datta 2015], preferentially showing ads for searching arrest records for visitors identified as African-American [Sweeny 2013], and differential pricing based on the physical location of a visitor [Hannack 2014].  These web profiles are also used to tailor search engine and news results to fit algorithmic predictions of what a user is predicted to be likely to read or view [Bozdag 2013, Boutin 2011, Hosanagar 2016].  The product of this algorithmic targeting is referred to as an "echo chamber" or "filter bubble" in which a slightly different version of the web is displayed for each visitor profile [Hannak et al.  2017, DiFonzo 2011, Pariser 2011, Gross 2011, Delaney 2017, Baer 2016].

These internet browsing behavior based filter bubbles are implicated as influencers of elections [Baer 2016, El-Bermawy 2016, Hern

2017, Jackson 2017], fake news [Spohr 2017; DiFranzo, and Gloria-Garcia 2017], and health decisions [Holone 2016;  Haim, Arendt and Scherr 2017].  Efforts to link internet user profiles to real world identities for enhanced targeting are underway.  Facebook is reported to have an effort underway to match electronic health record information with social media profiles to provide advertisers and healthcare professionals a broader view of an individual patient [Farr 2018, Farr 2017].

Coupling disease and treatment histories with an internet user profile could enable highly specific, disease stage based advertising and filtering.  Treatment options could be prioritized within the individuals filter bubble based on the advertising expenditures of pharmaceutical companies and even removed if the individual had characteristics deemed undesirable by the algorithm.  This potential is concerning, particularly when noting reports of discriminatory user profile targeting in other internet use cases and the central role advertising revenue plays in the financial success of technology companies such as Facebook and Google [Datta 2015, Sweeny 2013, Hannack 2014].

The known and potential impacts of web tracking technology on health information are significant.  Gaining a better understanding of the prevalence and characteristics of web tracking technology use on hospital websites is an important first step to understand the scope of this challenge.

This investigation is designed to determine the prevalence of web tracking technology use on the websites of Illinois hospitals.  The identity of the trackers will be analyzed to investigate usage trends and the data recipients.

## Methods

The Illinois Department of Public Health (IDPH) hospital directory was downloaded from the IDPH website (https://data.illinois.gov/dataset/410idph_hospital_directory).  This directory included all 210 licensed public hospitals in the state of Illinois as of September 27, 2017 and lists the hospital legal name, city and county the hospital is located in, and the type of hospital.

IDPH defined hospital types are:  general hospital, critical access hospital (no more than 25 inpatient beds), long-term acute care hospital (a hospital focused on patients who have an average hospital stay of more than 25 days), pediatric hospital (exclusively serves children), psychiatric hospital (a hospital focused on psychiatric care), and rehabilitation hospital (a hospital focused on rehabilitation after stabilization of acute medical issues).

For each hospital on the list, the Google search engine (https://www.google.com) was used to identify the hospital website by searching for the hospital name, city, and state.  The hospital web site was then visited using the Firefox browser (Version 52.5.0 for Microsoft Windows 7) with the Ghostery extension installed and activated (Version 8.0.7.1 by Cliqz International GmbH, available from https://ghostery.com).

Upon visiting each website, the Ghostery extension report on the number and identity of the web trackers found on the hospital web site were recorded into a spreadsheet for analysis.  These tests were performed on February 19, 2018.

The research protocol was reviewed by the Springfield Committee for Research Involving Human Subjects, and it was determined that this project did not fall under the purview of the IRB as research involving human subjects according to 45 CFR 46.101 and 45 CFR 46.102.

## Statistical analysis

Statistical analysis and graphs were produced using Microsoft Excel (Microsoft Office, 2013).

## Results

Websites could be identified for all 210 hospitals in the study sample. Analysis with the Ghostery tool revealed the presence of one or more web trackers on 197 (94%) hospital web sites (Table 1). Web trackers were present on all websites for pediatric and psychiatric hospitals. The average number of web trackers found on hospital web sites varied by hospital type and ranged from 2-4 trackers. Websites for general, pediatric, and long term acute care hospitals had the highest average number of web trackers. Critical access hospitals had the lowest average count of web trackers.

The most common web tracker used was Google Analytics, found on 156 (74%) Illinois hospital websites (Figure 1). Of the web trackers discovered, 88% were operated by Google and 26% by Facebook (Figure 2).

Table 1. Prevalence of web trackers on hospital websites in Illinois

| | Hospital Type | | | | | |
|---|---|---|---|---|---|---|
| | General | Critical Access | Pediatric | Psychiatric | LTAC | Rehab |
| | N = 133 | N = 51 | N = 3 | N = 10 | N = 9 | N = 4 |
| **Website with trackers (%)** | 131 (98%) | 42 (82%) | 3 (100%) | 10 (100%) | 7 (78%) | 4 (100%) |
| **Number of trackers (mean, SD)** | 3.5 (2.0) | 2.0 (1.7) | 4 (1.7) | 2.6 (1.9) | 4.1 (2.8) | 2.7 (2.1) |
| | | | | | | |
| **Web Tracker** | | | | | | |
|   **Google Analytics** | 87 (65%) | 34 (67%) | 2 (67%) | 6 (60%) | 7 (78%) | 2 (50%) |
|   **Google Tag Manager** | 64 (48%) | 8 (16%) | 2 (67%) | 3 (30%) | 5 (56%) | 1 (25%) |
|   **DoubleClick** | 35 (26%) | 1 (2%) | 2 (67%) | 1 (10%) | 1 (11%) | 0 |
|   **AddThis** | 34 (26%) | 6 (12%) | 0 | 2 (20%) | 1 (11%) | 0 |
|   **SiteImprove** | 27 (20%) | 5 (10%) | 0 | 1 (10%) | 6 (67%) | 0 |
|   **NewRelic** | 30 (23%) | 3 (6%) | 0 | 2 (20%) | 1 (11%) | 0 |
|   **CrazyEgg** | 23 (17%) | 3 (6%) | 1 (33%) | 1 (10%) | 0 | 0 |
|   **Facebook Pixel** | 23 (17%) | 3 (6%) | 1 (33%) | 0 | 1 (11%) | 0 |
|   **Google Translate** | 16 (12%) | 6 (12%) | 0 | 2 (20%) | 0 | 1 (25%) |
|   **Facebook Connect** | 7 (5%) | 12 (23%) | 1 (33%) | 0 | 0 | 2 (50%) |
|   **TradeDesk** | 12 (9%) | 3 (6%) | 0 | 0 | 0 | 0 |
|   **TypeKit** | 7 (5%) | 3 (6%) | 0 | 1 (10%) | 0 | 1 (25%) |
|   **ShareThis** | 6 (5%) | 2 (4%) | 0 | 2 (20%) | 0 | 1 (25%) |

Figure 1. Prevalence of web trackers by name on the websites of hospitals located in Illinois

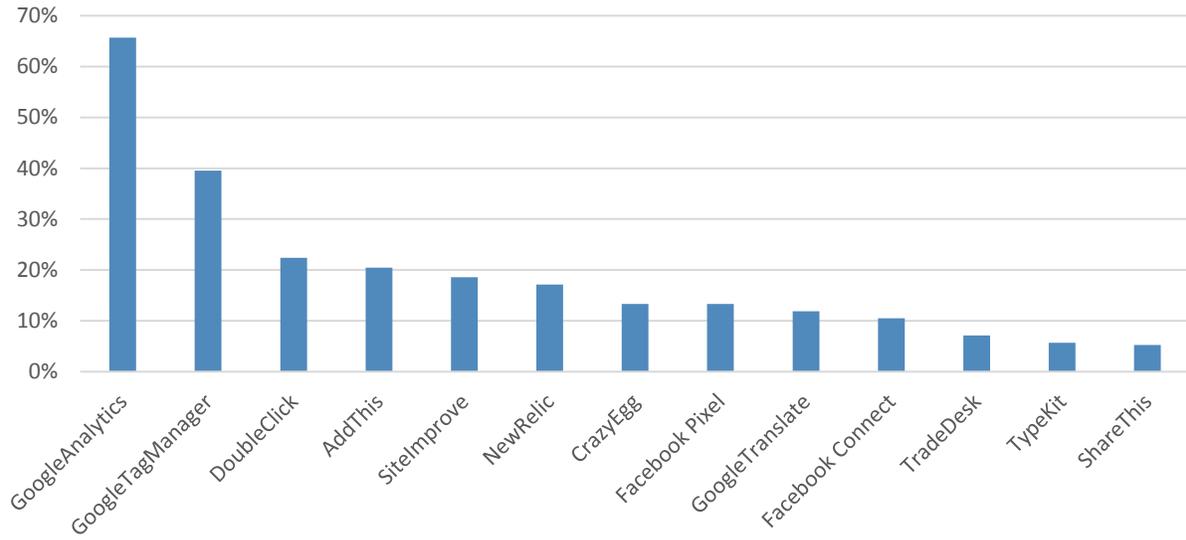

Figure 2. Prevalence of web tracker operator on the websites of hospitals located in Illinois

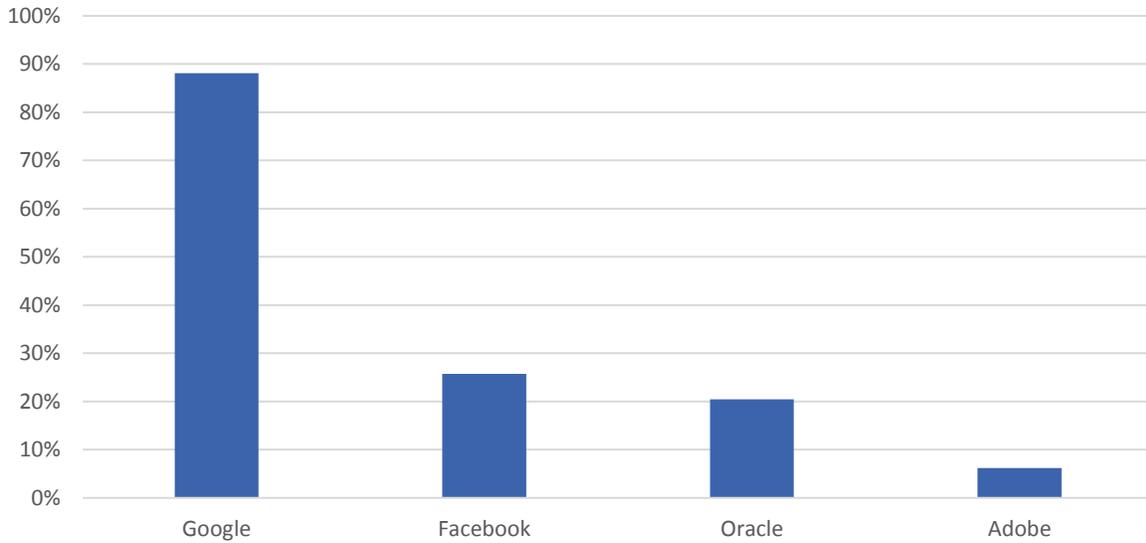

## Discussion

This analysis shows a high prevalence of web tracking technology (94%) on the websites of Illinois hospitals.  Most hospital websites employ 2 or more web tracking technologies.  The vast majority (88%) of Illinois hospital websites use web tracking technology operated by Google, Inc.  The closest competitor to Google is Facebook with tracking technology installed on 26% of hospital websites.  Little is known about the national prevalence of this technology on hospital websites.

The prevalence of Google and Facebook operated tracking technologies on the top one million websites based on traffic worldwide is similar [Englehardt and Narayanan, 2016].  Google Analytics, the most common tracker found in this study, is used on 65% of the top one million websites [BuiltWith 2018a].  Facebook Pixel can be found on 12% of the top one million websites [BuiltWith 2018b].  These rates are similar to what is seen on the websites of hospitals in Illinois.

The high prevalence of tracking technologies on the websites of Illinois hospitals give the hospitals using the technology, tracker operators, and marketers who may purchase this data unparalleled insight into the healthcare concerns of the people of Illinois.  Web tracking technology is useful to hospitals and hospital systems for the improvement of websites and monitoring the impact of advertising efforts.   Web tracker operators, such as Google and Facebook, reap considerable rewards by being able to target advertisements and tailor search results for website visitors based on detailed profiles created with this information.  These profiles are also marketed to other companies who wish to develop media campaigns to target individuals with specific characteristics.  Web site visitors benefit from the use of this technology in the form of free services such as Facebook and Google Gmail that are supported by an advertising ecosystem fueled by web tracker data.

In light of revelations about how web browsing profiles have been used and misused, search bubbles, and the potential for algorithmic discrimination hospital leadership and policy makers must carefully consider if it is appropriate to use third party tracking technology on hospital web sites.

## Conclusions

Web tracking technology, predominately in the form of Google services, is extremely common on the websites of public hospitals in Illinois.

Further research is required to determine broader trends in web tracking technology use on hospital websites and how this technology use may impact the public.